\documentclass[aps,prl,twocolumn,superscriptaddress,fleqn,a4paper]{revtex4-1}
\usepackage[ngerman,english]{babel}
\usepackage[utf8]{inputenc}   
\usepackage{xspace}
\usepackage{amsmath,tensor}
\usepackage{array,color}
\usepackage{relsize,exscale}  
\usepackage{textcase}	
\usepackage{soul,color}		
\usepackage{calc} 
\usepackage{makeidx,showidx}
\usepackage{graphicx}
\usepackage{wrapfig}
\usepackage{float}
\usepackage{textcomp}
\usepackage{siunitx}
\usepackage[version=3]{mhchem}
\usepackage{enumitem}
\usepackage{fancyref}
\usepackage[hidelinks=true]{hyperref}
\sethlcolor{green}
\setulcolor{red}
\usepackage{lineno}			

\newcommand{\erww} [1] {\ensuremath{\langle {#1} \rangle}}

\newcommand{\lsco} {{La$_{2-x}$Sr$_x$CuO$_4$}\@\xspace}

\newcommand{\hgryb} {{HgBa$_{2}$CuO$_{4+\delta}$}\@\xspace}

\newcommand{\ybco} {$\ce{YBa2Cu3O_{6+y}}$\@\xspace}

\newcommand{\ybcoE} {$\ce{YBa2Cu4O8}$\@\xspace}

\newcommand{\tc} {\ensuremath{T_{\rm c}}\@\xspace}
\newcommand{\cperp}{\ensuremath{c \bot B_0}\@\xspace}
\newcommand{\cpara}{\ensuremath{{c\parallel\xspace B_0}}\@\xspace}

\newcommand{\beq} {\begin{equation}}
\newcommand{\eeq} {\end{equation}}

\newcounter{exex}[section]

\makeatletter\newcommand\listofexamples{\section*{List of Examples}\@starttoc{xmp}}
	\newcommand\l@example[2]{\par\noindent#1~\textit{#2}\par}
\makeatother






\makeatletter
\renewcommand\subsection{\@startsection 
{subsection}{3}{0mm}
{-\baselineskip}
{0.5\baselineskip}
{\centering \textbf }}
\makeatother

\makeatletter
\renewcommand\subsubsection{\@startsection 
{subsubsection}{3}{0mm}
{-\baselineskip}
{0.5\baselineskip}
{\centering  }}
\makeatother

\makeindex

\begin{document}
\title{NMR shift and relaxation and the electronic spin of superconducting cuprates}
\author{Marija Avramovska}
\author{Danica Pavi\'cevi\'c}
\author{Jürgen Haase}
\affiliation{University of Leipzig, Felix Bloch Institute for Solid State Physics, Linn\'estr. 5, 04103 Leipzig, Germany}

\begin{abstract}
Very recently, there has been significant progress with establishing a common phenomenology of the superconducting cuprates in terms of nuclear magnetic resonance (NMR) shift and relaxation. Different from the old interpretation, it was shown that the shifts demand two coupled spin components with different temperature dependencies. One spin component couples isotropically to the planar Cu nucleus and is likely to reside at planar O, while the other, anisotropic component has its origin in the planar copper $3d(x^2-y^2)$ orbital. Nuclear relaxation, on the other hand, was found to be rather ubiquitous and Fermi liquid-like for planar Cu, i.e., it is independent of doping and material, apart from the sudden drop at the superconducting transition temperature, $T_{\rm c}$. However, there is a doping and material dependent anisotropy that is independent on temperature, above and below $T_{\rm c}$. Here we present a slightly different analysis of the shifts that fits all planar Cu shift data. In addition we are able to derive a simple model that explains nuclear relaxation based on these two spin components. In particular, the only outlier so far, \lsco, can be understood, as well. While this concerns predominantly planar Cu, it is argued that the two component model should fit all cuprate shift and relaxation data.
\end{abstract}

\maketitle
\vspace{0.5cm}
{\centering {\today}}
\labelformat{paragraph}{#1}

\subsection{Introduction}
Nuclear magnetic resonance (NMR) is a powerful local, bulk probe of material properties \cite{Slichter1990}. This concerns the chemical as well as electronic structure of materials, which can be studied locally at various nuclear sites in the unit cell. The changes in the NMR shifts and relaxation from the modification of the density of states due to the opening of a superconducting gap in conventional superconductors are famous examples \cite{Hebel1957,Yosida1958}. 

Not surprisingly, after the discovery of cuprate superconductivity \cite{Bednorz1986} NMR experiments focused in particular on planar Cu and O in these type-II materials (for reviews of cuprate NMR see \cite{Slichter2007,Walstedt2008}). However, with the early focus on a few systems and a missing guidance from an established theory, the NMR data interpretation ceased to evolve as a whole.

Over the last decade, it was shown with various NMR experiments on \ce{La{_{1.85}}Sr{_{0.15}}CuO4} \cite{Haase2009}, \ybcoE \cite{Meissner2011}, and samples of the \hgryb family of materials \cite{Haase2012,Rybicki2015} that one of the cornerstones of the old interpretation was in fact not correct. A single temperature dependent spin component, $s(T)$, that follows from the uniform spin susceptibility, i.e., $s(T) = \chi(T) \cdot B_0$, in an external field $B_0$, is not able to describe the temperature dependent NMR spin shifts, $\tensor[^n]{K}{_\alpha}(T)={^{n}H}_{d}\cdot \chi(T)$. The shifts can be measured at any nucleus, $n$, or for any orientation ($d$) of the external field with respect to the crystal axes, with the (anisotropic) hyperfine constants, ${^{n}H}_{d}$, describing the interaction between nuclear and electronic spin. 

In the most simple extension of the model, a two-component description was introduced \cite{Haase2009}, which has two spin components that couple with two different hyperfine coefficients, $\tensor[^{n}]{H}{_{1d}}$ and $\tensor[^{n}]{H}{_{2d}}$, to each nuclear spin, $n$. On general grounds, two spin susceptibilities ($\chi_1, \chi_2$) demand a third term from a coupling between the two electronic spin components. That is, one has to write $\chi_1 = \chi_{11}+\chi_{12}, \chi_{2}=\chi_{22}+\chi_{21}$ ($\chi_{12}=\chi_{21}$), and,
\begin{equation}\label{eq:2comp}
\tensor[^n]{K}{_\parallel_,_\perp} = {^{n}H}_{1\parallel,\perp} \cdot(a+c) + {^{n}H}_{2\parallel,\perp} \cdot(b+c)
\end{equation}
with $a=\chi_{11} B_0$, $b=\chi_{22} B_0$, and $c = \chi_{12} B_0$, and the magnetic field parallel and perpendicular to the crystal $c$-axis.

With this expression for the shifts, the above mentioned experiments could be explained, with uncertainties arising only from the unknown orbital shift contributions \cite{Rybicki2015}. Note that it is notoriously difficult to separate the temperature independent orbital shift term, $\tensor[^{n}]{K}{_L_\parallel_,_\perp}$, from the total magnetic shift, $\tensor[^{n}]{\hat{K}}{_\parallel_,_\perp} (T)  = \tensor[^{n}]{K}{_L_\parallel_,_\perp} + \tensor[^n]{K}{_\parallel_,_\perp} $, since also the spin shift can be temperature independent (e.g. from the Pauli susceptibility of a regular Fermi liquid). In particular for planar Cu where one expects large orbital shifts, this leads to high uncertainties. Until recently \cite{Slichter2007,Walstedt2008,Rybicki2015}, it was assumed that the residual low temperature NMR shift is the orbital shift,
\beq\label{eq:orbital}
\tensor[^n]{\hat{K}}{_\parallel_,_\perp} (T \rightarrow 0) =: \tensor[^n]{K}{_L_\parallel_,_\perp},
\eeq
since spin singlet pairing should lead to the disappearance of the spin term as one approaches the lowest temperatures, i.e. $\tensor[^n]{K}{_\parallel_,_\perp}(T \rightarrow 0) \approx 0$, to a good approximation.

However, as was pointed out in early experiments \cite{Pennington1989} and discussed later as well \cite{Renold2003}, the thus defined and measured orbital shift for \cpara, $\tensor[^{63}]{K}{_L_\parallel}= \tensor[^{63}]{\hat{K}}{_\parallel} (T \rightarrow 0)$, is too large for this direction of the magnetic field (not for \cperp). In addition, for \ybco or \lsco there is no significant temperature dependence of the magnetic shift when the field is parallel to the $c$-axis, i.e. $\tensor[^{63}]{\hat{K}}{_\parallel}$ is temperature independent also at the lowest temperatures (apart from uncertainties with respect to latent diamagnetism in the mixed state \cite{Barrett1990b}), while $\tensor[^{63}]{\hat{K}}{_\perp}$ is strongly temperature dependent. In the single spin component view this was explained by assuming $\tensor[^{63}]{H}{_\parallel}\equiv A_\parallel+4B \approx 0$, where $A_{\parallel,\perp}$ is the anisotropic hyperfine coefficient expected for a spin in the Cu $3d(x^2-y^2)$ orbital, while $B$ is an isotropic transferred term from the 4 neighboring Cu atoms in a single band picture.
\begin{figure}
\centering
\includegraphics[width=0.45\textwidth ]{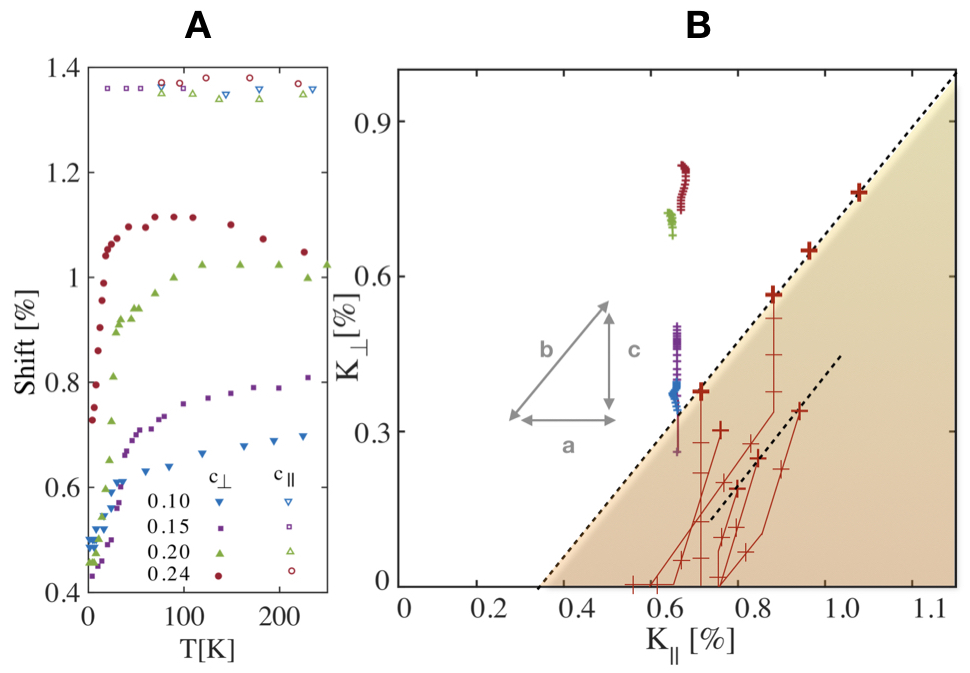}
\caption{(A) Total $^{63}$Cu shifts vs. temperature, $\hat{K}_{\parallel,\perp}(T)$, for 4 doping levels of \lsco and two directions (\cpara, \cperp) of the external field $B_0$ with respect to the crystal $c$-axis (adopted from \cite{Ohsugi1994}). $K_\parallel$ is $T$ independent and similar for all doping levels. $K_\perp$ shows a much larger spread with doping and decreases rapidly near \tc.  (B) Sketch of the spin shift $K_\perp(T)$ vs $K_\parallel(T)$ plot valid for all cuprates \cite{Haase2017}, with real data from the 4 doping levels of \lsco (with temperature as an implicit parameter). The shaded area is where the rest of the many cuprates can be found (the shaded triangle has a hypotenuse of slope $\approx 1$), only typical data are shown by crosses. Data lie on straight line segments (lines) with a few slopes only: a slope $ \approx 1$ (dashed lines); a very steep slope (vertical lines); a slope of 2.5. For example a slope of 2.5 is typical for \hgryb (at higher $T$), a steep slope for \ybcoE, a slope of 1 for some Tl-based compounds, as well as for overdoped \hgryb at low $T$. \lsco is a clear outlier with only the steep slope. In the simple two component description, cf.~\eqref{eq:2para2}, \eqref{eq:2perp2}, a change in one of the components $a, b$ or the coupling $c$ leads to the indicated slopes in the middle of (B).}\label{fig:fig1}
\end{figure}

Very recently, by plotting all magnetic shift data available in the literature \cite{Haase2017}, a mere inspection of the total shifts, $\tensor[^{63}]{\hat{K}}{_\parallel_,_\perp}(T)$, revealed that a single component view is indeed not possible. This can easily be seen since the shift variations as a function of temperature are only proportional to each other in certain intervals of temperature (or doping), for which one finds a few different slopes $\kappa = \Delta \tensor[^{63}]{K}{_\perp}/ \Delta \tensor[^{63}]{K}{_\parallel}$, indicative of special relationships.

This more or less generic behavior of all cuprates was observed for all systems, with the exception of the family of \lsco \cite{Haase2017}, cf.~Fig.~\ref{fig:fig1}. 

\begin{figure}[t]
\centering
\includegraphics[width=0.45\textwidth ]{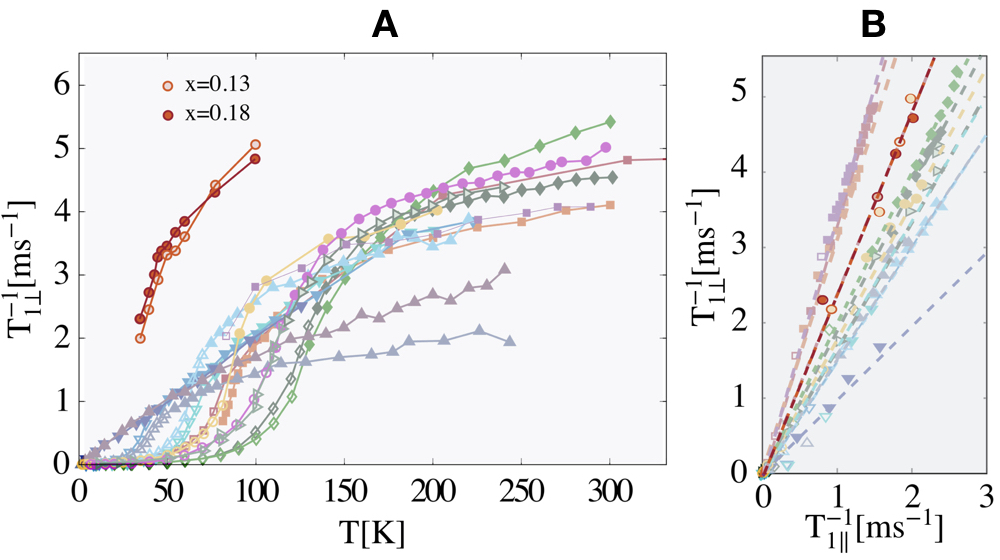}
\caption{(A) Planar $^{63}$Cu relaxation rates of the cuprates (data from \cite{Jurkutat2019}); $1/T_{1\perp}$ of \lsco in comparison is about twice as high as that of other cuprates. (B) $1/T_{1\perp}$ vs $1/T_{1\parallel}$, which is $\approx 2.3$ for \lsco (highlighted), is very similar to what is found for other cuprates (data \cite{Jurkutat2019}). It is mostly $1/T_{1\parallel}$ that changes with doping and material, but remains proportional to $1/T_{1\perp}$ (at all temperatures)}.\label{fig:fig2}
\end{figure}

Another important, very recent observation followed from gathering and plotting all available literature Cu relaxation data \cite{Avramovska2019,Jurkutat2019}. Surprisingly, one finds generic behavior, as well, with the exception of just one family, \lsco. All the other cuprates have rather similar relaxation rates, $1/T_{1\perp}$, i.e. if measured with the magnetic field perpendicular to the crystal $c$-axis, \cperp, cf.~Fig.~\ref{fig:fig2}. In particular, one finds that just above \tc the value of $1/T_{1\perp}T_{\rm c}\sim \SI{20}{/Ks}$ is very similar for all cuprates, while \tc can be very different, or even close to zero for strongly overdoped systems. This is an indication of ubiquitous Fermi liquid-like behavior. There is no particular doping dependence of $1/T_{1\perp}$ as one might naively expect if electronic spin fluctuations beyond those of a more regular Fermi liquid were to increase towards lower doping levels (there are hardly data available at very low doping). In fact, a value of \SI{20}{/Ks} follows with the Korringa relation \cite{Korringa1950} from those cupates with the highest shifts, i.e. the upper right corner of the shaded triangle in Fig.~\ref{fig:fig1}, which suggests \cite{Avramovska2019} that the shift have the tendency to be suppressed if the Korringa relation fails, and it is not due to an increased relaxation rate, but a suppressed shift. At higher temperatures, $1/T_{1\perp}$ begins to lack behind a Fermi liquid's $1/T_1 \propto T$ behavior.  Below \tc, $1/T_{1\perp}T$ is universal if plotted over the reduced temperature $T/T_{\rm c}$. 

The situation is somewhat different for $1/T_{1\parallel}$ since for this direction of measurement (\cpara) the rates differ between families and have the tendency to increase with decreasing doping. However, it was demonstrated that the ratio, $\left[1/T_{1\perp}(T)\right]/\left[1/T_{1\parallel}(T)\right]$, is  temperature independent for all cuprates, and is the same above and below \tc, i.e. both rates are proportional to each other \cite{Avramovska2019,Jurkutat2019}. For the highest doped cuprates, nearly isotropic relaxation is found, and the ratio increases as the doping decreases and shows a maximum of about 3.3 for \ybcoE (where data are available). 

Interestingly, the \lsco family of materials is the only outlier to this phenomenology, cf.~Fig.~\ref{fig:fig2}. However, the anisotropy ratio is also temperature independent and has a value of about 2.3, very similar to that of other cuprates.


\subsection{Planar Cu Shifts}

The total magnetic shift for planar Cu, $\tensor[^{63}]{\hat{K}}{_\parallel_,_\perp}$, is the sum of an orbital and spin shift component, and we have for the two orientations (\cpara, \cperp) of the magnetic field $B_0$ with respect to the crystal $c$-axis,
\begin{equation}
\tensor[]{\hat{K}}{_\parallel_,_\perp}(T) = \tensor[]{K}{_L_\parallel_,_\perp}+\tensor[]{K}{_\parallel_,_\perp}(T).\label{eq:twoorb1}
\end{equation}
It is of particular use to plot the total shifts $\tensor[]{\hat{K}}{_\perp}(T)$ vs. $\tensor[]{\hat{K}}{_\perp}(T)$ \cite{Haase2017} with temperature as an implicit parameter, i.e. one does not make assumptions about $\tensor[]{K}{_L_\parallel_,_\perp}$. Such a plot brings out a number of remarkable trends \cite{Haase2017,Avramovska2019}. A sketch of such a plot is presented in Fig.~\ref{fig:fig1} (B), and we repeat some conclusions \cite{Avramovska2019}, but also include new ones, below.

A fundamental assumption is \cite{Haase2017,Avramovska2019},
\beq\label{eq:orbitalperp}
\tensor[]{K}{_{L\perp}} \approx 0.30\%,
\eeq
since all cuprates show a rather similar low temperature shift for \cperp with $\tensor[]{\hat{K}}{_\perp}(T\rightarrow 0) \approx 0.30\%$. Therefore, this value appears to be reliable. It is the same assumption made early on \cite{Slichter2007}. We note that this value is  backed by first principle calculations \cite{Renold2003} (while this is not the case for $\tensor[]{K}{_{L\parallel}}$).

\begin{figure}[]
\centering
\includegraphics[width=0.3\textwidth ]{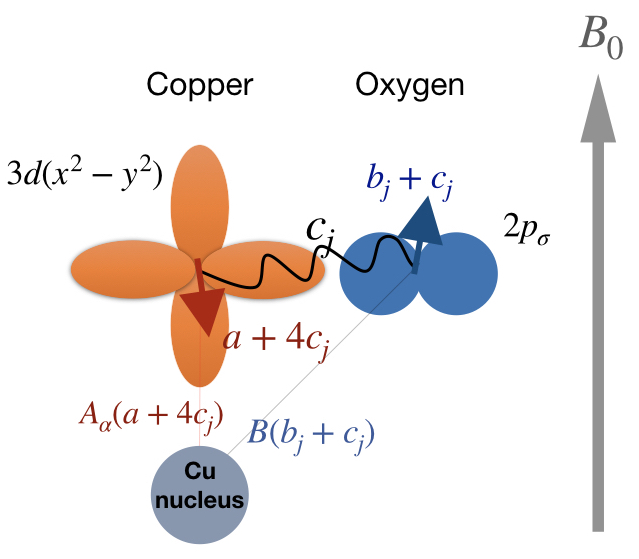}
\caption{In an external magnetic field $B_0$ two spin components $a$ and $b_j$ appear, originating from the planar Cu $3d(x^2-y^2)$ and the four surrounding O $2p_\sigma$ orbitals, respectively. Due to a coupling ($c$) the effective components are $(a+4c_j)$ and $(b_j+c_j)$. While $(b_j+c_j)$ is positive for the cuprates, ($a+4c_j$) turns out to be negative. The hyperfine coefficients $A_\alpha$ and $B$ lead to orientation dependent ($\alpha$) NMR shifts $\tensor[]{K}{_\alpha}=A_\alpha (a+4c_j) + B (b_j+c_j)$ at the Cu nucleus.} \label{fig:fig3}
\end{figure}
Second, except for \lsco, all data points in that plot are found in the lower right triangle that has as hypothenuse a line of slope 1, i.e. $\Delta \tensor[]{\hat{K}}{_\perp}/\Delta \tensor[]{\hat{K}}{_\parallel}(T) \approx 1$. This line points immediately to an isotropic hyperfine coefficient, while the fact that $\tensor[]{\hat{K}}{_\parallel}(T) > \tensor[]{\hat{K}}{_\perp}(T)$ (all data in the lower, right triangle) demands a second, very anisotropic hyperfine coefficient that acts mostly for \cpara. Very similar arguments as put forward in the old literature let us choose $A_{\perp,\parallel}$ and $B$: note that there must be spin in the $3d(x^2-y^2)$ orbital, and it is very likely that there will also be an isotropic coupling term. Then, the NMR shifts demand, however, that the spin polarization in the $3d(x^2-y^2)$ orbital must be \emph{negative}, as pointed out recently \cite{Haase2017,Avramovska2019}, for we know that $A_\parallel$ is negative, and $|A_\parallel |\gg A_\perp $ \cite{Husser2000}.

Thus, we write with \eqref{eq:2comp},
\begin{equation}\label{eq:2compCu}
\tensor[]{K}{_\parallel_\perp} = A_{\parallel,\perp} \cdot (a+4c_j) + B \cdot 4(b_j+c_j).
\end{equation}
For symmetry reasons we take $(b_j+c_j)$ from each of the 4 neighbors to be the same, i.e. from spin in the planar O 2$p_\sigma$ orbitals, cf.~Fig.~\ref{fig:fig3}.
As before \cite{Avramovska2019}, we will neglect $A_\perp$ and simply write,
\begin{gather}
\tensor[]{K}{_\parallel} = A_{\parallel} (a+4c_j) + B \cdot 4(b_j+c_j)\label{eq:2para2}\\
\tensor[]{K}{_\perp} \approx B \cdot4(b_j+c_j). \label{eq:2perp2}
\end{gather}
This is a different notation from before \cite{Avramovska2019} where we used $b = 4 b_j$. 

A zero spin shift, our first, fundamental assumption means that $\sum_j(b_j+c_j) = 0$, and we have for the other orientation,
\beq\label{eq:intersection}
\tensor[]{\hat{K}}{_\parallel}(T \rightarrow 0) = \tensor[]{{K}}{_L_\parallel} + A_\parallel (a+4c_j). 
\eeq
In order to estimate the orbital shift, $\tensor[]{{K}}{_L_\parallel}$, for this orientation of the field, as before \cite{Avramovska2019}, the most reliable approach is to use \eqref{eq:orbitalperp} together with calculations of the orbital shift anisotropy, since the latter is mostly determined by matrix elements involving the orbital bonding wave functions of Cu and O \cite{Pennington1989,Renold2003}. In fact, we use the suggested value of 2.4 from \cite{Renold2003},
\beq\label{eq:orbparallel}
\tensor[^{63}]{K}{_{L\parallel}} = 2.4\cdot \tensor[^{63}]{{K}}{_L_\perp} \approx 0.72\%. 
\eeq
Note that this value could vary between families, but since the orbital shift for \cperp does not change significantly between families, we do not expect a large effect for \cpara, as well. This is important as it means that most cuprates have a non-vanishing spin shift for \cpara from a negative spin polarization in the $3d(x^2-y^2)$ orbital, even at the lowest tempeartures.

As mentioned earlier, a few special slopes govern the shift-shift plot presented in their figure 7 \cite{Haase2017}, and we highlighted them in Fig.~\ref{fig:fig2}(B), again. These are segments defined by temperature or doping for which the ratio of changes in both shifts is constant, ${\Delta \tensor[]{\hat{K}}{_\perp}(T)}/{\Delta \tensor[]{\hat{K}}{_\parallel}(T)} = \kappa,$
and one finds 4 slopes, $\kappa \approx 0, 1, 2.5, \infty$. For example,  $\kappa = 1$ denotes isotropic shift lines and readily follows from a mere change of $b_j$ only, as it enters both terms in \eqref{eq:2para2} and \eqref{eq:2perp2}. Then, $\kappa \approx 0$ in this approximation is realized by a change in $a$, only, since we neglected the rather small $A_\perp$. Note that term $c$ operates on both shifts, $\tensor[]{K}{_\perp}$ and $\tensor[]{K}{_\parallel}$, and must be involved in the special slopes $\kappa = 2.5$ and $\kappa \approx \infty$. While not favored before \cite{Avramovska2019}, we believe that $\kappa \approx \infty$ is caused by a mere change in $c$. The reasoning is as follows: not a single material in the shift-shift plot shows a negative slope, i.e. a slope to the right of $\kappa \approx \infty$. This is remarkable and must mean that the component $a$ cannot significantly be involved in shift changes. 

With this assumption that $c_j$ causes $\kappa \approx \infty$, we note that \eqref{eq:2para2} and \eqref{eq:2perp2} require, 
\beq\label{eq:cancellation}
A_\parallel \approx - B,
\eeq
and we have with \eqref{eq:2para2} and \eqref{eq:2perp2},
\begin{gather}
\tensor[]{K}{_\parallel} \approx B (4b_j-a)\label{eq:2para3}\\
\tensor[]{K}{_\perp} \approx B 4(b_j+c_j). \label{eq:2perp3}
\end{gather}
Note that in this approximation, $c$ effectively acts only for \cperp. Then, the slope of $\kappa \approx 2.5$ is given by a concomitant change of $b_j$ and $c_j$, e.g. $\Delta b_j = 1.5 \Delta c_j$ if both terms change proportionally.

To summarize, in the above model the individual changes of $a$, $b_j$, and $c_j$ correspond to slopes of $\kappa = 0, 1$, and $\infty$, respectively, in Fig.~\ref{fig:fig1} (if all $b_j$ and $c_j$ are the same). Even if this is not precisely what happens, we think that \eqref{eq:2para3} and \eqref{eq:2perp3} still capture the fundamental aspects of the planar Cu shifts.\par\medskip

With these results in mind we can look at the data for \lsco again. 


The high temperature shifts for \lsco, $\tensor[]{K}{_\perp}$, are much larger than what we expect from its $\tensor[]{K}{_\parallel} = B(4b_j-a)$ values. In one scenario, a larger $a$ and larger $b_j$ could position this family at larger $\tensor[]{K}{_\perp}$ (for given $c_j$). The action of a temperature dependent $c_j$ then leads to the $\kappa \approx \infty$ slope. Alternatively, $c_j$ could be  much larger for \lsco, i.e. much more positive, at high temperatures. This also leads to a much larger $B(b_j+c_j)$. Again, a drop in $c_j$ then makes $(b_j+c_j)$ disappear.  

To conclude, while \lsco is an outlier in the shifts, the position in Fig.~\ref{fig:fig1} can be understood within the two component scenario, as well.

\subsection{Planar Cu Relaxation}

The nuclear relaxation rate $1/T_{1\parallel}$ measures the in-plane fluctuating magnetic fields, $\erww{\tensor*{h}{*_\perp^2}}$, from electronic spin fluctuations, while $1/T_{1\perp}$ is affected by both, in-plane, $\erww{\tensor*{h}{*_\perp^2}}$, as well as out-of-plane, $\erww{\tensor*{h}{*_\parallel^2}}$, fields (only fluctuating field components perpendicular to the nuclear quantization axis lead to nuclear spin flips, required for spin-lattice relaxation).

Phonons will cause nuclear relaxation for quadrupolar nuclei ($I>1/2$, like Cu and O) as they modulate the electric field gradient, but it has been shown that the magnetic fluctuations dominate in most situations \cite{Takigawa1991b,Suter2000b}, and the recent analysis of all Cu relaxation data shows that a simple magnetic mechanism appears to capture the overall behavior quite well \cite{Avramovska2019,Jurkutat2019}. 

In a straightforward approach one would assume nearly isotropic spin fluctuations filtered by the nuclear hyperfine coefficients, which can then lead to a relaxation anisotropy. The electronic correlation time ($\tau_0$) of electronic spin fluctuations is expected to be very fast compared to the slow precession of the nuclei. Thus, the nuclear relaxation rates can be written as \cite{Pennington1989},
\begin{gather}
\frac{1}{T_{1\parallel}} = \frac{3}{2} {\gamma^2} \cdot 2 \erww{\tensor*{h}{*_\perp^2}} \tau_0 \label{eq:t1para}\\
\frac{1}{T_{1\perp}} = \frac{3}{2} {\gamma^2} \left[ \erww{\tensor*{h}{*_\perp^2}}+ \erww{\tensor*{h}{*_\parallel^2}}\right]\tau_0,\label{eq:t1perp}
\end{gather}
from which the relaxation anisotropy follows,
\begin{gather}
\frac{1/T_{1\perp}}{1/T_{1\parallel}} = \frac{1}{2} +\frac{\erww{\tensor*{h}{*_\parallel^2}}}{2\erww{\tensor*{h}{*_\perp^2}}}.\label{eq:ratiot1}
\end{gather}

Given that the shifts demand two different electronic spin components coupled to the nuclei through an anisotropic constant $A_{\parallel,\perp}$ and an isotropic constant $B$, one should allow for two different fluctuating spin densities $\alpha$ and $\beta =  \sum_j \beta_j$, as well. Furthermore, since  the fluctuations are caused by rapid exchange, the correlation time $\tau_0$ should be the same for both components.

We thus write,
\begin{gather}
\erww{\tensor*{h}{*_\perp_,_\parallel^2}} \approx \erww{(\sum_jB\beta_j+A_{\perp,\parallel}\alpha)^2}\label{eq:field1}\\
\erww{\tensor*{h}{*_\parallel^2}} \approx B^2\erww{(\sum_j\beta_j-\alpha)^2},\label{eq:hpara}\\
\erww{\tensor*{h}{*_\perp^2}} \approx B^2\erww{(\sum_j\beta_j+f\alpha)^2},\label{eq:hperp}
\end{gather}
where we introduced $f=A_\perp/B \approx -A_\perp/A_\parallel$ if $A_\perp \alpha$ is not negligible (see below).
\begin{figure}[t]
\centering
\includegraphics[width=0.4\textwidth ]{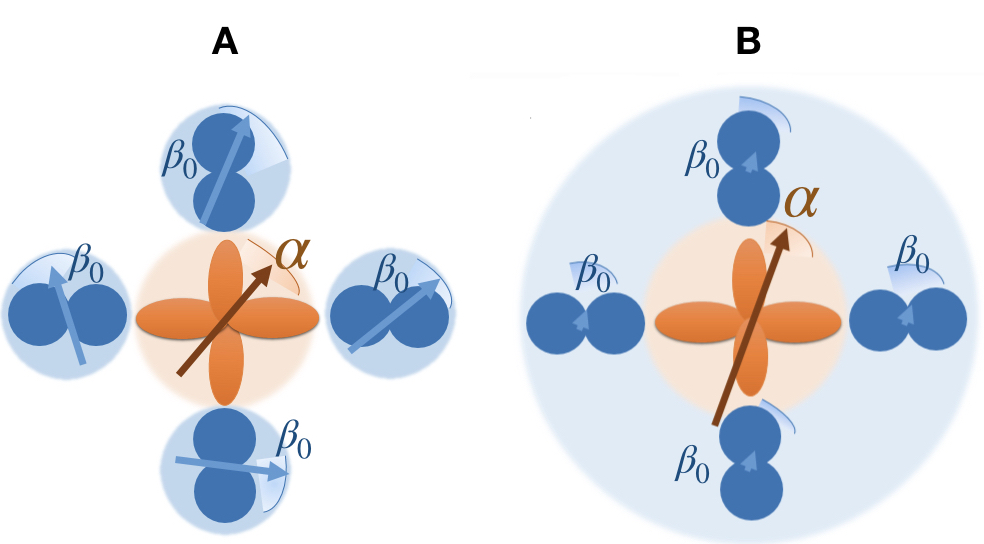}
\caption{Fluctuating spins $\alpha$ and $\beta_0$, respectively located in the Cu $3d(x^2-y^2)$ and O 2$p_\sigma$ orbital. (A), all 5 spin components fluctuate independently, i.e., $\erww{\alpha \beta_0}=0, \erww{\beta_i\beta_j}= \beta_0^2 \delta_{ij}$, (B), the fluctuations are fully correlated, i.e., $\erww{\alpha \beta_0}=\alpha \beta_0, \erww{\beta_i\beta_j}= \beta_0^2$.}\label{fig:fig4}
\end{figure}

With these expressions for the fluctuating field components we seek to explain a rather doping and material independent $1/T_{1\perp}$ (it only increases marginally with decreasing doping) and a material and doping dependent $1/T_{1\parallel}$ that explains the temperature independent anisotropy \eqref{eq:ratiot1}, as well as the exceptional behavior found for \lsco.

In a first scenario one might be interested to see what would be the consequences of totally uncorrelated spin fluctuations for the 5 spin components, i.e. $\erww{\beta_i\beta_j} = \erww{\beta_0^2}\delta_{ij}$, and $\erww{\beta_j\alpha} = 0$, cf.~Fig.~\ref{fig:fig3}. We then have $\erww{h_\perp^2} = 4 \erww{\beta_0^2}$ and $\erww{h_\parallel^2}=4 \erww{\beta_0^2}+\erww{\alpha^2}$, thus with \eqref{eq:t1para} and \eqref{eq:t1perp} for uncorrelated ($u$) fluctuations,
\begin{gather}
\frac{1}{T_{1\parallel,u}} = \frac{3}{2} {\gamma^2}  B^2\cdot 8 \erww{\beta_0^2} \tau_0 \label{eq:t1para2}\\
\frac{1}{T_{1\perp, u}} = \frac{3}{2} {\gamma^2} B^2 \cdot\left[ 8\erww{\beta_0^2}+ \erww{\alpha^2}\right]\tau_0,\label{eq:t1perp2}
\end{gather}
and if follows for the anisotropy,
\beq
\frac{1/T_{1\perp,u}}{1/T_{1\parallel,u}} = 1+\frac{\erww{\alpha^2}}{8\erww{\beta_0^2}}\label{eq:ratiow1}.
\eeq
Clearly, for $\erww{\alpha^2} \lesssim \erww{\beta_0^2}$ we find near isotropic relaxation, and in order to explain the largest anisotropy of about 3.3 \cite{Jurkutat2019}, we conclude  $\erww{\alpha^2} \approx 18.4 \erww{\beta_0^2}$. This implies, however, rather large changes of $\alpha$ and $\beta$ for meeting the experimental observations, i.e. the change in relaxation between materials and different doping levels, which appears to be difficult to meet in this approach (we do notice that a large $\alpha$ could be present, which demands that we do not neglect $A_\perp$ for the modeling of nuclear relaxataion).\par\medskip

\begin{figure}[t]
\centering
\includegraphics[width=0.3\textwidth ]{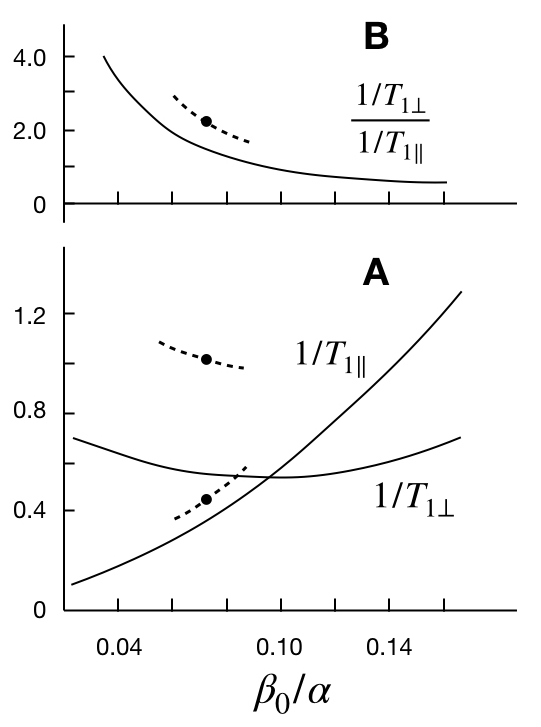}
\caption{(A) Calculated nuclear relaxation rates for \cpara ($1/T_{1\parallel}$) and \cperp ($1/T_{1\perp}$) as a function of the ratio of the two spin components $\beta_0 \equiv \beta_j$ and $\alpha$ ($\beta_0/\alpha$), according to \eqref{eq:t1para}, \eqref{eq:t1perp} with \eqref{eq:hpara} and \eqref{eq:hperp}, in arbitrary units. (B) The anisotropy of the relaxation \eqref{eq:ratiot1} varies between 4 and 0.5 in the same range of $\beta_0/\alpha$. Corresponding line segments for \lsco with an anisotropy of about 2.3 are indicated, as well.}\label{fig:fig5}
\end{figure}

In a second scenario we assume that all spins are aligned, i.e. the 5 fluctuating spin components are correlated, with $\erww{\beta_i\beta_j} = \erww{\beta_0^2}$ and $\erww{\beta_j\alpha}=\pm \alpha \beta_0$. We note that the field fluctuations $\erww{\tensor*{h}{*_\perp^2}} \approx B^2\erww{(\sum_j\beta_j+f\alpha)^2}$ that enter \eqref{eq:t1para} and  $\erww{\tensor*{h}{*_\parallel^2}} \approx B^2\erww{(\sum_j\beta_j-\alpha)^2}$ that determine \eqref{eq:t1perp} are both quadratic in the resulting local spin densities. Therefore, in order to find a rather flat dependence for the relaxation for \cperp on $\beta_0$, as demanded by the experiment, we need to be close to its minimum, while at the same time, the parabola must be shifted by a negative $\alpha$ compared to the other parabola in order to meet a smaller but varying relaxation rate for \cpara. The results of simple calculations according to \eqref{eq:t1para}, \eqref{eq:t1perp} with \eqref{eq:hpara} and \eqref{eq:hperp} are shown in Fig.~\ref{fig:fig5}. 
We observe that there is only a special region with solutions that fit the experiments, for ${\beta_0}/\alpha = 0.04$ to $0.11$ according to anisotropies ranging from $3.3$ to $ 1.0$, respectively, cf. Fig.~\ref{fig:fig5}.

Furthermore, an increase of $\alpha$ by a factor of about 1.3, at an anisotropy ratio of 2.3, increases the relaxation rates in both directions by about a factor of two, cf.~Fig.~\ref{fig:fig5}, which readily explains the data found for \lsco. We thus conclude that the two component $\beta_0$ and $\alpha$ are crucial for the cuprates, but appear to be very similar for most of the materials.

\subsection{Discussion}

It seems out of question that a two-component scenario describes the shifts and relaxation in the cuprates quite well. It has spin density located in the Cu $3d(x^2-y^2)$ orbital, which couples to the nucleus through the rather anisotropic hyperfine constant $A_{\parallel,\perp}$, and, most likely, the planar O $2p_\sigma$ orbital, leading to an isotropic hyperfine interaction given by~\eqref{eq:2compCu}. 

The spin density $\alpha$ is much larger than $\beta_0$, as one expects from the overall material properties, however, the uniform response of both spins is quite different, also due to the coupling term $c_j$. 

The special slopes observed in the shift-shift plot, cf.~Fig.~\ref{fig:fig1}, are caused by changes of the individual spin components as function of doping or temperature, except for the slope $\kappa \approx 2.5$ that must stem from a concomitant change of $b_j$ and $c_j$. This leads to the simple conclusion that $A_\parallel \approx - B$ (while $A_\perp \approx 0.15 A_\parallel$ \cite{Husser2000}), and it leaves us with a straightforward description of the spin shifts of the cuprates in terms of \eqref{eq:2para3} and \eqref{eq:2perp3}, i.e., $\tensor[]{K}{_\parallel} \approx B (4b_j-a)$ and $\tensor[]{K}{_\perp} \approx B 4(b_j+c_j)$ (in these equations we also adopted a different notation in terms of $b_j$ compared to our earlier analysis \cite{Avramovska2019}). 
We note that the conclusion that $A_\parallel \approx -B$ has a similar origin as in the old interpretation.

Looking again at Fig.~\ref{fig:fig1} (B), the cuprates are sorted in this shift-shift plot effectively by the high-temperature $b_j$, the component that grows with increasing doping (towards the upper right in Fig.~\ref{fig:fig1}(B)). As the temperature is lowered, at a given temperature, which can be above or at \tc, this term begins to disappear due to the action of $c_j$ (both components $c_j$ and $b_j$ can fall together, as well). It is the coupling to $a$ that sets $c_j$ (and effectively couples different $a$ terms, as well). The component $a$ appears to be temperature independent. It emerges that either $b_j$ or $c_j$ can be exhausted independently indicated by changes in slope at lower temperatures. However, all cuprates seem to reach the same $(b_j+c_j)=0$, which we define as zero spin shift. Importantly, there is no evidence that there is a different mechanism as one passes through \tc if the shift began to change already far above \tc (NMR pseudogap), but $c_j$ can traverse the region below \tc at a much higher rate for given steps in temperature.

The earlier conclusion \cite{Haase2017,Avramovska2019} that the spin shift for \cpara does not disappear at low temperatures, here takes the formulation that $(4b_j - a)\neq 0$ and says that the positive spin density $b_j$ and the negative spin density $a_j$ can remain temperature independent for systems with $\kappa \approx \infty$, or $b_j$ can also drop with $c_j$ as for the systems with slope $\kappa \approx 2.5$, i.e. it does not change in the condensed state, while relaxation ceases.

In terms of a simple fluctuating field model we can explain the cuprate relaxation rather well. Fast electronic, Fermi liquid-like spin fluctuations act through two different hyperfine coefficients with two different electronic spin densities on the Cu nucleus (or, these densities are part of that ubiquitous fluid). The corresponding fluctuations from the 5 locations must be correlated, as one might have guessed due to the close proximity. The onsite $3d(x^2-y^2)$ spin ($\alpha$) is about 10 times as large as that due to one O neighbor $(\beta_0)$. The spin density $\alpha$ appears to be the same for all cuprates, except for the \lsco family it is 30\% larger. The spin $\beta_0$ varies with doping and between materials and leads to the change in $1/T_{1\parallel}$ observed in the data. For large doping the relaxation anisotropy that is about 1 and it increases to about 3.3 for \ybcoE (corresponding to a change in $\beta_0$ of about 3). For \lsco the anisotropy is 2.3 and thus also $\beta_0$ is about a factor of two larger. 

Since the $^{63}$Cu relaxation begins to disappear only at \tc, for all cuprates, the electronic, Fermi liquid-like spin fluctuations freeze out and the relaxation disappears. Thus, the pseudogap in the relaxation is just due to the correlations that for planar Cu do not change $\alpha$ and $\beta_0$ spin alignment. For planar O the situation is different as the nucleus couples to two $\alpha$ spins at adjacent Cu nuclei and their coupling changes, which leads to the pseudogap in the relaxation for nuclei that are affected by different $a$ spins \cite{Avramovska2020b}.

The relation between the spin densities $\alpha, \beta_0$ and the uniform response of the system in terms of $a$, $b_j$, and $c_j$ is not known. It appears that the response of $\alpha$ is rather small compared to that of $\beta$, which may not be surprising since different $a$ should favor antiferromagnetic alignment (that is somehow affected by $\beta$).

It appears that the doping dependent spread in $\tensor[]{{K}}{_\parallel}$ varies among the cuprates. This reminds us of the way the charge carriers enter the \ce{CuO2} plane \cite{Jurkutat2014}. For the \lsco family, the doped charges $x$ enter almost exclusively the $2p_\sigma$ orbital ($n_p$) while for other systems the Cu $3d(x^2-y^2)$ ($n_d$) is affected as well ($x = \Delta n_d + 2 \Delta n_p$ \cite{Haase2004,Jurkutat2014}), and the spread in doping appears to grow with $\Delta n_d$. The maximum achievable \tc, however, is set by the sharing of the parent material's hole content, i.e., $n_d^* + 2n_p^* =1$ and 
$T_{\rm c, max} \propto n_p^*$ \cite{Jurkutat2014,Rybicki2016,Jurkutat2019b}. 
Materials with the highest \tc appear to adopt $\kappa \approx 2.5$, only. However, the jumping between different slopes $\kappa$ in different regions of the shift-shift plot that involves $b_j$ and/or $c_j$ below \tc is absent for optimally doped systems, which probably means that $b_j$ and $c_j$ are matched at optimal doping. 

Finally, one may argue that the intra cell charge variation between neighboring planar O atoms that appears to be ubiquitous and that can respond to the external magnetic field \cite{Reichardt2018} could be involved in the two component scenario.

\subsection{Conclusions}
Two spin densities were shown to reside in the planar Cu $3d(x^2-y^2)$ and likely the planar O $2p_\sigma$ orbitals, respectively, with hyperfine constants $A_{\parallel,\perp}$ and $B \approx - A_\parallel$. They connect the Cu nuclear spins with a rather ubiquitous Fermi liquid-like bath. The relaxation anisotropy is predominantly due to changes in the planar O spin density that increases with doping. Near \tc these electronic fluctuations freeze out and the relaxation disappears. 

The uniform response $a$ and $b_j$ of the two electronic spins on Cu and O is special in the sense that $a$ is negative while $b_j$ points along the field. The coupling term $c_j$ between $a$ and $b_j$ sets the temperature dependence of the shift above (NMR pseudogap) and below \tc. Interestingly, at the lowest temperatures $4(b_j+c_j)$ approaches the same value for all cuprates, probably zero, but $a$ remains and most of $b_j$, as well, resulting in a non-vanishing spin shift for \cpara, $K_\parallel \approx (4b_j - a) \neq 0$. 

The coupling term $c_j$ must be related to a coupling between different spin components $a_i$ on different Cu nuclei, and it is argued that the pseudogap phenomenon for planar O nuclear relaxation and that of Y is just a consequence of the temperature dependence of $c_j$, an effect that cannot be there in the Cu relaxation data.

This simple two-component scenario appears to fit all cuprates, in particular also the only outlier family so far, \lsco, which must make it a reliable framework for theory.

\subsection*{Acknowledgements}
We acknowledge support from Leipzig University, and fruitful discussions with A. Poeppl, M. Jurkutat, A. Kreisel.

\subsubsection*{Author contributions}
J.H. introduced the main concepts and held the overall leadership; all authors were involved in data analysis and discussion equally, as well as in preparing the manuscript.

\vspace{0.5cm}

\bibliography{JHGedanken.bib}   

\printindex
\end{document}